\documentclass[aps,prd,twocolumn,reprint,superscriptaddress,nofootinbib,preprintnumbers,longbibliography]{revtex4-1}
\usepackage[english]{babel}
\usepackage[utf8]{inputenc}
\usepackage{amssymb}
\usepackage{amsmath}
\usepackage{color}
\usepackage{hyperref}
\usepackage{bbm}
\usepackage{epsfig}
\usepackage{multirow}
\usepackage[percent]{overpic}

\usepackage[normalem]{ulem}

\usepackage[dvipsnames]{xcolor}
\usepackage{xspace}

\newcommand{\ex}{\Delta\varepsilon_{\rm ex}}

\newcommand{\s}{\sigma}

\newcommand{\up}{\uparrow}
\newcommand{\down}{\downarrow}
\newcommand{\w}{\omega}

\newcommand{\de}{{\rm d}}
\newcommand{\dd}{\partial}

\newcommand{\Spin}{\mathcal{S}}
\newcommand{\dub}[1]{\mathcal{D}^{#1}}

\newcommand{\T}{\mathcal{T}}
\newcommand{\la}{\langle}
\newcommand{\ra}{\rangle}
\newcommand{\ket}[1]{\left| {#1} \right\rangle }
\newcommand{\bra}[1]{\left\langle {#1} \right| }
\newcommand{\mean}[1]{\left\langle {#1} \right\rangle }
\newcommand{\GF}[1]{\langle\!\langle #1\rangle\!\rangle}

\renewcommand{\Im}{\mathrm{Im}}
\newcommand{\dk}{d^\dagger}
\newcommand{\ck}{c^\dagger}
\newcommand{\entropy}{\mathbb{S}}
\newcommand{\Simp}{\mathbb{S}^{\rm imp}}

\newcommand{\Sec}[1]{Sec.~\ref{sec:#1}}
\newcommand{\beq}{ \begin{equation} } 
\newcommand{\eeq}{ \end{equation} }
\newcommand{\beqa}{\begin{eqnarray}}
\newcommand{\eeqa}{\end{eqnarray}}
\newcommand{\nn}{\nonumber}
\newcommand{\es}{& = &}

\newcommand{\fig}[1]{Fig.~\ref{fig:#1}}
\newcommand{\figs}[1]{Figs.~\ref{fig:#1}}
\newcommand{\eq}[1]{Eq.~(\ref{#1})}

\newcommand{\ie}{\textit{i.e.~}}

\hypersetup{
    bookmarks=false,        
    colorlinks=true,        
    linkcolor=red,        
    citecolor=blue,        
    filecolor=blue,      
    urlcolor=blue
}

\begin{document}

\title{Magnetic Kondo regimes in a frustrated half-filled trimer}
	   
\author{Krzysztof P. W{\'o}jcik}
\email{kpwojcik@ifmpan.poznan.pl}
\affiliation{Institute of Molecular Physics, Polish Academy of Sciences, 
			 Smoluchowskiego 17, 60-179 Pozna{\'n}, Poland}
\affiliation{Physikalisches Institut, Universit\"{a}t Bonn, Nussallee 12, D-53115 Bonn, Germany}

\author{Ireneusz Weymann}
\affiliation{Faculty of Physics, A.~Mickiewicz University, 
			 Uniwersytetu Pozna\'{n}skiego 2, 61-614 Pozna{\'n}, Poland}

\author{Johann Kroha}
\affiliation{Physikalisches Institut, Universit\"{a}t Bonn, Nussallee 12, D-53115 Bonn, Germany}

\date{\today}

\begin{abstract}
We analyze theoretically the phase diagram of a triangular triple quantum dot
with strong onsite repulsion coupled to ferromagnetic leads. 
This model includes the competition of magnetic ordering of local or itinerant 
magnetic moments, geometric frustration and Kondo screening.  
We identify all the phases resulting from this competition.
We find that three Kondo phases -- the conventional one, the two-stage
underscreened one, and the one resulting from the ferromagnetic Kondo 
effect -- can be realized at zero temperature, and all are very susceptible 
to the proximity of ferromagnetic leads. In particular, we find that 
the quantum dots are spin-polarized in each of these phases. 
Further, we discuss the fate of the phases at non-zero temperatures,
where a plethora of competing energy scales gives rise to complex landscape
of crossovers. Each Kondo regime splits into a pair of phases, one not 
magnetized and one comprising magnetically polarized quantum dots.
We discuss our results in the context of heavy-fermion physics in 
frustrated Kondo lattices.
\end{abstract}

\maketitle

\section{Introduction}
\label{sec:intro}

Heavy-fermion systems are magnetic materials where rare-earth magnetic 
ions reside on a lattice, and their $4f$ electrons carrying local moments
hybridize with the itinerant electrons of a conduction 
band \cite{Loehneysen07}. The resulting spin exchange coupling between local 
and itinerant moments leads to a Kondo effect and, hence, a heavy band near 
the Fermi level \cite{Loehneysen07,Kirchner20}. 
The competition between local spin exchange and non-local spin coupling 
mediated by the RKKY interaction \cite{RK,K,Y} can induce a quantum phase 
transition (QPT) from a paramagnetic Kondo-screened phase with 
expanded Fermi volume to a magnetically ordered phase \cite{Loehneysen07}.
However, experiments indicate that in some heavy-fermion 
systems the ordered and the Kondo phases may be separated by a state 
which is neither long-range ordered nor completely Kondo-screened \cite{Steglich2009,Paschen2010May,Kuchler2017Dec}. 
This suggests that in the global heavy-fermion phase diagram magnetic 
frustration may play an additional, important role \cite{Si2006,Vojta2008Sep,Coleman2010Oct,Si2014May,Grube2018}. The latter may be induced by the 
long-range, oscillatory nature of the RKKY interaction.   
Frustration in insulating spin lattices has been largely treated on the 
basis the two-dimensional Shastry-Sutherland model 
\cite{ShastrySutherland81,Sachdev01}. However, the presence of a conduction
band with potential Kondo singlet formation introduces another complication. 
Despite analytical \cite{Burdin2002Jul,Coleman2010Oct} and numerical 
\cite{Motome2010Jul,Assaad2018Mar} treatments of frustrated Kondo lattice 
models a complete understanding of all the phases possible by multiple 
tuning parameters is still lacking. The problem becomes even more complex 
in systems where the local magnetic moments sit on several,
crystallographically inequivalent lattice sites
\cite{VojtaIneq,Ineq2010,Ineq2010b,Ineq2011,Ineq2013,Ineq2016} 
or where magnetic order may even coexist with a Kondo-screened 
phase \cite{KondoFMcoex}.

\begin{figure}[t!]
\centering
\includegraphics[width=0.95\columnwidth]{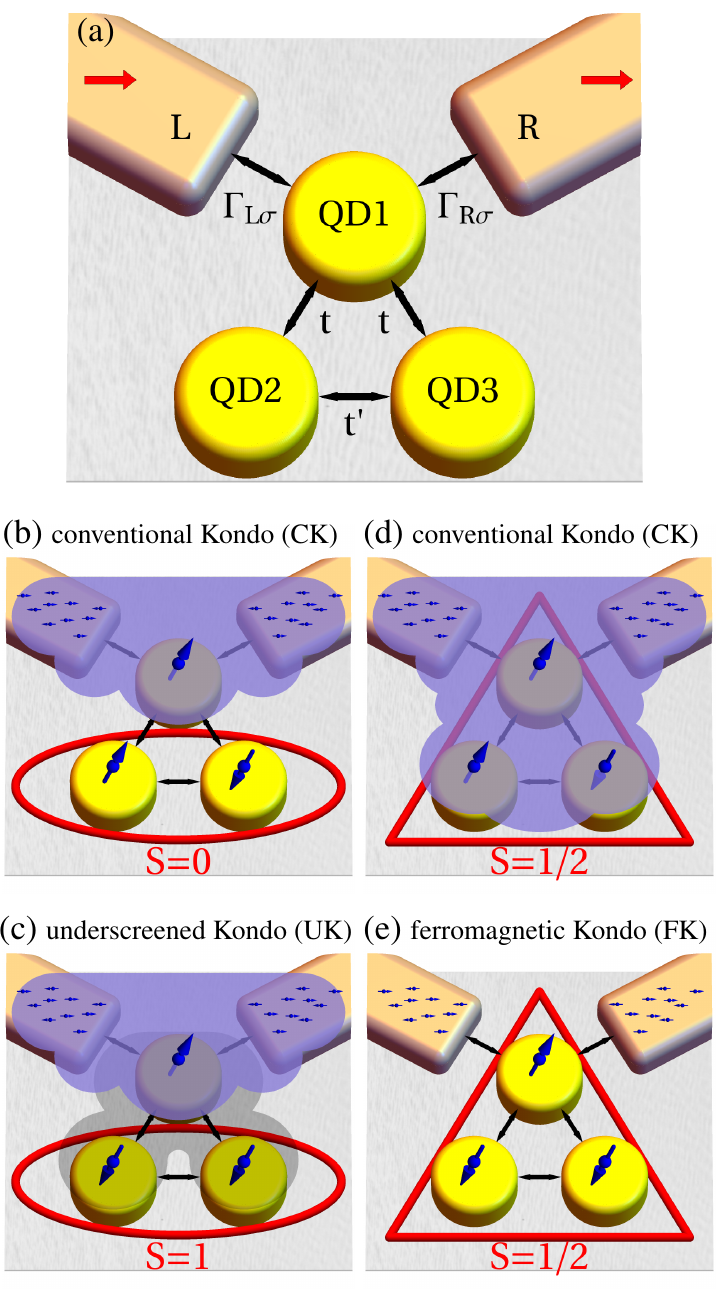}
\caption{
		 (a) Schematic of the considered Kondo trimer, 
		 consisting of three \emph{quantum dots}, denoted as QD1, QD2, QD3,
		 coupled to ferromagnetic left (L) and right (R) leads 
		 with spin-dependent coupling strengths $\Gamma_{{\rm L(R)}\s}$.
		 (b)-(e) Illustrative presentation of possible orientations of the quantum dot spins
	     (illustrated as arrows) and phases that may arise in the system:
	       (b) the \emph{conventional Kondo} (CK) phase with weak inter-dot hopping~$t$,
	       (c) the \emph{underscreened Kondo} (UK) phase,
	       (d) the CK phase for stronger $t$ and
	       (e) the \emph{ferromagnetic Kondo} (FK) phase.
	       $S$ denotes the spin of the indicated fragment of the trimer system.
	       The semi-transparent shapes represent schematically the Kondo screening clouds.
	     See \Sec{qualitative} for more details.
		 }
\label{fig:system}
\end{figure}

In the present work, to shed some more light on the intriguing
interplay of the aforementioned effects, we 
take a minimal quantum impurity model, where 
frustration, Kondo screening and magnetic correlations coexist
and compete with each other. 
In particular, we consider a quantum-dot (QD) trimer coupled to a single 
spin-polarized screening channel, in the geometry 
depicted in \fig{system}(a).
The three quantum dots, exhibiting strong on-site 
Coulomb interactions, are assumed to form a triangular constellation.
The first quantum dot (QD$1$) is embedded between two leads made of a ferromagnetic metal.
This dot is coupled to the second (QD2) and third (QD3) quantum dot,
respectively, via the hopping matrix element $t$,
while QD$2$ and QD$3$ are coupled by the  frustrating hopping $t'$.
This model incorporates the essential features
of the interplay of local Kondo screening, magnetic ordering (magnetic 
dimer formation), and geometric frustration, parameterized by the ratio
$t'/t$. It also takes into account the inequivalence of Kondo 
sites in that only the first dot is coupled to the leads. This is a simplified, 
numerically tractable description of a situation where a spatially 
varying density of states or exchange coupling may lead to an 
exponential suppression of the Kondo temperature on some of the 
Kondo sites, i.e., an effective decoupling of 
some screening channels \cite{Mitchelldmft,BullaPseudoGap,AndersNatPhys}.
A possible coexistence of itinerant 
magnetic ordering and Kondo screening can be analyzed by allowing for 
a magnetic polarization of the leads. Such quantum impurity 
model has such advantage that, despite its complex physics, it can be 
reliably analyzed by the numerical renormalization group (NRG) method \cite{WilsonNRG},
and  that it can be realized in quantum dot experiments where all the system parameters 
can be continuously tuned, which is usually difficult in lattice systems. 
Note also that a quantum impurity model of this type would emerge in a 
cluster dynamical mean-field theory \cite{DMFT,DMFTbook} of 
geometrically frustrated Kondo or Anderson lattice systems.

Non-magnetic Kondo trimers have been extensively examined theoretically
\cite{Zitko2008Jun,Vernek2009Jul,Numata2009Oct,Oguri2011May,Koga2012Nov,3qdReview,
Mitchell_QPT,Mitchell_phase_diag,Lopez2013Jan,Tooski2014Jun,Park3qdChargeKondo,
Tooski2016Jan,Koga2016May,Niklas2017Mar,Sun3QD,3QD-pseudogap},
also in the case of few-channel screening 
\cite{Mitchell2010Feb,ZitkoBoncaPRL,3QD-3CH,Aligia2018,Konig2020Feb}.
The presence of a QPT
separating the \emph{conventional Kondo} (CK) phase \cite{Kondo,bookHewson}
from the exotic \emph{ferromagnetic Kondo} (FK) regime 
hosting a non-screened local magnetic moment and 
singular dynamics at low temperatures \cite{poorMan,singularFK}
is well-established \cite{Mitchell_QPT}. 
A separation of the \emph{underscreened Kondo} (UK) phase 
\cite{NozieresBlandin,singularFK}
from the FK one by zero-temperature crossover has 
been also analyzed \cite{Mitchell_phase_diag}. 
In addition, quantum dot trimers have also been widely studied in the context of quantum computing
\cite{Loss2003Apr,Kostyrko2009Feb,Busl2010Mar,Weymann2011May,Bulka2011Jan,
Hsieh2012Sep,Taylor2013Jul,Seo2013Jan,Luczak2014Oct}
and spintronics
\cite{Kuzmenko2006Feb,Korkusinski2007Mar,Emary2007Dec,IW_3qd_2015,IW_3qd_2018,Wrzesniewski2020Apr}.
However, especially the latter remains quite detached from the studies 
of strongly-correlated Kondo physics.
In particular, a comprehensive analysis of all the phases 
in the presence of magnetic order seems to be missing. 
One of the aims of this paper is therefore to fill this gap.
We show that due to the ferromagnetic proximity effect
all the Kondo phases (namely CK, FK and UK), turn into their 
spin-polarized counterparts (which we denote by CK', FK' and UK',
correspondingly) for arbitrarily small frustrating  coupling $t'$. 

While in the existing literature the trimer phase diagrams have been
investigated mainly at vanishing temperature,  
we show that the $T\to 0$ limit is  irrelevant for experiments in certain 
parameter regimes. 
Moreover, even though often the quantum impurity systems 
can generally be understood in terms of a few stable $T\to 0$ 
phases and the QPTs between them
\cite{Vojta,Vojta2,ZitkoQPTs,Andrei_TIAM,Galpin2005May,Zitko2008Jun,Baruselli,
AndersNatPhys}, a number of cases, where continuous crossovers significantly 
alter the physics, are also known \cite{KMWWb,ZitkoCrossovers,Zitko3QDs,Bulla}, 
especially in the context of competition between the Kondo
effect and the spin polarization caused by ferromagnetic 
leads \cite{MartinekEx,KWIW-P,KWIW-2Kondo}. 

We demonstrate that in the presence of frustration the Kondo phases
are spin-polarized in the $T\to 0$ limit and remain so 
up to experimentally relevant temperatures even for very 
weak frustrating coupling. 
This means that our results are actually relevant also 
for nearly linear trimers with $t'$ interpreted as a weak 
next-nearest-neighbor hopping; cf. \fig{system}(a). 
Additionally, a finite-temperature
crossover links the corresponding spin-polarized and spin-isotropic phases.
As can be expected \cite{IW_UKpol}, the underscreened Kondo (UK') phase is especially 
fragile to the presence of magnetic leads, which tangibly
differs it from the ferromagnetic Kondo (FK') phase, where the spin polarization 
of relevant quantum dots is significantly smaller. This is in contrast 
to huge resemblance between the non-magnetic UK and FK regimes 
\cite{Mitchell_phase_diag}.

It is also important to note that a number of experiments on triple quantum dot systems
have been performed in the context of quantum computing 
\cite{exp1,Vidan2004Oct,exp2,Rogge2008May,Gaudreau2009Aug,Granger2010Aug,
	Laird2010Aug,3QDexp1,Braakman2013Apr}
or charge frustrations \cite{Seo2013Jan},
and to study the triple QD's Kondo physics \cite{Oguri2011May}.
We therefore believe that our results will foster further experimental efforts
to examine different competing phenomena in correlated magnetic nanostructures,
especially in coupled quantum dot systems.

As the study concerns quite a rich model, to increase its 
accessibility we describe the results beginning with the most 
general features, and providing more precise understanding in 
subsequent sections.
We start from presenting a qualitative physical picture 
of the system in \Sec{qualitative}, setting up the stage for 
the landscape of phases emerging in the presence of spin-polarized 
leads. Then, having presented the details 
of the model and methodology in \Sec{model}, we list and estimate 
all the relevant energy scales in \Sec{E}. The general
structure of the phase diagram in the space of inter-dot hopping
$t$, frustrating hopping $t'$ and the temperature is outlined
in \Sec{PD}. Finally, the numerical results allowing to precisely 
pinpoint the borders between the phases are presented in \Sec{NRG},
corroborating estimations done in \Sec{E}. 
The paper is concluded in \Sec{conclusions}.

\section{Qualitative physical picture}
\label{sec:qualitative}

The system depicted in \fig{system}(a) comprises several couplings,
each of them driving some kind of spin ordering
even in the absence of leads magnetization.
In the present section we therefore briefly
summarize all known Kondo phases of such system 
\cite{Mitchell_QPT,Mitchell_phase_diag}, 
in a form suitable for the comparison with our results presented in further sections.

First of all, the possibility of electron hopping between the first
and other dots, described by hopping amplitude $t$,
tends to align the spin of the first dot antiparallelly
with respect to the spins of the other dots (QD2 and QD3).
On the other hand, the direct hopping between QD2 and QD3, $t'$,
frustrates this order by trying to bind the electrons
residing the second and third dot into a spin singlet.
This becomes possible, when the QD1 spin 
gets screened by the leads via the (conventional) Kondo effect. 
Such scenario is presented schematically in \fig{system}(b).

The situation is completely different in the case of relatively small $t'$. 
Then the QD2-QD3 direct exchange interaction is weak, and can 
be overcome by the super-exchange mediated by QD1. The latter 
is always ferromagnetic in sign \cite{Mitchell_phase_diag} and 
can bind QD2-QD3 subsystem into a spin triplet. However, no matter 
how small $t$ is, it always causes the anti-ferromagnetic exchange
coupling of the resulting spin with the Fermi liquid formed by
the Kondo screening at QD1. Therefore, at sufficiently cryogenic 
conditions further Kondo screening occurs, which, with only one 
screening channel, results in the underscreened Kondo effect in the
QD2-QD3 subsystem and spin-doublet ground state for the whole 
system. This is illustrated in \fig{system}(c), with CK screening
cloud represented as in \fig{system}(b), and the more transparent 
shape illustrating the cloud corresponding to partial screening.

The picture presented so far is based on the assumption that 
the Kondo coupling of the first dot to the leads is sufficiently strong
such that the inter-dot interactions do not give rise to the formation of 
molecular-like trimer states at temperatures above the Kondo
temperature. However, with increasing $t$, the state depicted in 
\fig{system}(b) continuously evolves into the one schematically 
represented in \fig{system}(d). The latter no longer contains 
practically decoupled QD2-QD3 singlet. Instead, $t$ correlates
QD2 and QD3 with QD1, spreading the spin doublet from QD1 into 
the whole trimer. Nevertheless, for strong enough $t'$,
the tendency of anti-parallel alignment guarantees that the trimer 
ground state would remain a spin doublet, which can be efficiently
screened at sufficiently low temperatures by conventional Kondo correlations.

Finally, for weak frustrating coupling $t'$ and strong hopping $t$,
the trimer acquires a magnetically staggered structure, with QD1 spin aligned 
antiparallelly to the spins of QD2 and QD3, which leaves 
the trimer in spin doublet state. However, unlike for 
the strong $t'$ case, now the coupling of the trimer to the 
leads is \emph{ferromagnetic}, as the trimer spin direction 
follows the alignment of spins of QD2 and QD3 and is opposite 
to the direction of QD1 spin. In this way the ferromagnetic Kondo system 
is formed, as sketched in \fig{system}(e). 

As shown in the following, all these phases are vulnerable
to the symmetry breaking introduced by the spin polarizations of the leads $p$.
At non-zero temperatures each phase splits into two regimes,
separated by a crossover, with one comprising 
partially spin-polarized trimer stable in the low-temperature 
limit, and the other comprising non-magnetized trimer, 
which is relevant only at elevated temperatures.


\section{Model and methods}
\label{sec:model}

The trimer coupled to the leads is modeled by a Hamiltonian 
of the general form $H=H_{\rm L} + H_{\rm R} + H_{\rm T} + H_{\rm 3QD}$, 
where the left (L) and right (R) leads are described by a single effective
band \cite{even-odd},
$H_{\rm L}+H_{\rm R} = \sum_\s \int \w \ck_{\w\s}c^{}_{\w\s} \de\w $, 
with $\ck_{\w\s}$ denoting the creation operator for an electron of energy $\w$ 
and spin $\s$ in a combination of relevant wave functions in respective 
electrodes coupled to the trimer. An effective hybridization is given by
$\Gamma_\s(\w) = \Gamma_{{\rm L}\s}(\w)+\Gamma_{{\rm R}\s}(\w)$ \cite{even-odd},
so that the tunneling Hamiltonian reads
\beq
H_{\rm T} = \sum_{\s} \int  \sqrt{\frac{\Gamma_{\s}(\w)}{\pi}} (\ck_{\w\s}d^{}_{1\s} + {\rm H.c.})\de\w ,
\eeq
with $\dk_{i\s}$ creating a spin-$\s$ electron in QD$i$, $i=1,\,2,\,3$. 
Note that only QD$1$ is coupled to the leads, cf. \fig{system}(a). 
We assume constant hybridization functions within
the band of width $2D$, $\Gamma_{r\s}(\w) = \Gamma_{r\s}(0)$ for $|\w|<D$
($\w=0$ at the Fermi energy), with sharp cutoff at energies $\pm D$. 
For the subsequent calculations, the magnetization of the leads 
(assumed parallel in the two leads) is represented by spin-dependent, 
left-right symmetric effective couplings, 
$\Gamma_{L\s}=\Gamma_{R\s} = \Gamma_{\s}=(1+\s p)\Gamma/2$, 
with $\Gamma$ measuring the coupling strength and  $p$ denoting the effective
spin-polarization of the leads \cite{MartinekEx}. 
Assuming equal onsite repulsion $U$ on each QD, the trimer Hamiltonian is written as 
\beqa
H_{\rm 3QD} \es \sum_{i\s} \left(-\frac{U}{2} + \delta_i \right)n_{i\s} 
				+ \sum_i U n_{i\up}n_{i\down}
			\nn\label{HQD}\\
				&+& \sum_{i,j,\s} t_{ij}\dk_{i\s}d^{}_{j\s},
\label{H3QD}
\eeqa
where the summations run over $i,j \in \{1,\,2,\,3\}$, but $i\neq j$, $t_{ij}=t_{ji}$, and $\delta_i$
denotes the detuning of QD$i$ from local particle-hole symmetry (PHS) point. 
The hoppings to two side-coupled QDs are assumed equal, 
$t_{12}=t_{13}=t$, while the frustrating coupling $t'=t_{23}$ is kept independent; see also \fig{system}(a). 

To analyze the properties of this system,
we use the numerical renormalization group 
technique \cite{WilsonNRG}. We construct
the full density-matrix from states discarded during the calculation
\cite{AndersSchiller1,AndersSchiller2,Weichselbaum}, 
and use an open-access code \cite{fnrg} as a basis for our implementation.
This method allows us to reliably capture 
the full spectrum of the discretized Hamiltonian
and calculate all the physical quantities directly from the spectral data,
without any need for artificial broadening \cite{adaptive}.
In particular, the linear conductance through the system at temperature $T$ is calculated from \cite{MeirWingreen}
\beq
G = \frac{e^2}{h} \sum_\s \int \!\!\left[ -\frac{\dd f(\w)}{\dd \w}\right] \T_{\s}(\w) \de\w  \; ,
\label{MWformula}
\eeq
where $f(\w)$ is the Fermi-Dirac distribution, and the spin-resolved 
transmission coefficient $\T_{\s}(\w)$ is given in terms of the retarded
Green's function of the first quantum dot as
\beq
\T_{\s}(\w) = -\Gamma_\s \Im \GF{\dk_{1\s} | d^{}_{1\s}}(\w).
\label{TC}
\eeq
The latter can be obtained in Lehmann representation directly from the 
NRG solution. The expectation values of the operators defined within
the trimer subspace are obtained by taking the trace with the relevant
reduced density matrices \cite{fnrg,fnrgPaper}.

To fully understand the nature of different Kondo phases,
we also calculate the trimer contribution to the entropy.
It can be found from 
\beq
\Simp = \entropy^{\rm tot} - \entropy^{0}, 
\label{SimpDef}
\eeq
where $\entropy^{\rm tot}$ denotes the entropy of the full system, whereas $\entropy^{0}$
is the entropy of the system without the trimer.
The entropy is calculated directly from the 
spectrum of the discretized Hamiltonian
(for a given set of parameters, one additional calculation is needed to determine $\entropy^{0}$).

Throughout the paper we use the band cutoff as the energy unit, $D=1$,
and take the onsite repulsion $U$ equal to the bandwidth, $U=D$, unless 
stated otherwise. The temperature is expressed in units of 
energy, \ie the Boltzmann constant $k_B\equiv 1$.
The leads spin polarization is assumed to be, $p=0.5$,
yet the $p=0$ case is also considered for comparison.
The system is assumed to be at the local PHS point, 
$\delta_i = 0$; cf.~\eq{H3QD}.  
In the NRG calculations, we take the coupling strength to be $\Gamma=U/10$, 
the discretization parameter $\Lambda=3$, and the number of states kept 
at each iteration is $N=3000$.

\section{Relevant energy scales}
\label{sec:E}

The most important low-temperature phases have been outlined in the discussion of \fig{system}.
In the present section we elaborate on them further, precisely explaining their origin.
To determine the remaining phases, the phase boundaries between them and their fate at elevated temperatures, 
we discuss the relevant energy scales, in particular the exchange field $\ex$ induced by the ferromagnetic 
leads. The results presenting the trimer's ground state and the exchange field 
as a function of hoppings $t$ and $t'$ are shown in Figs.~\ref{fig:2a} and \ref{fig:ex}. 
Then, the quantitative phase diagram is discussed in \Sec{PD}
and presented in \fig{PD}.

\subsection{Isolated trimer}
\label{sec:3QD}

\begin{figure}[tb]
	\centering
	\includegraphics[width=0.8\linewidth]{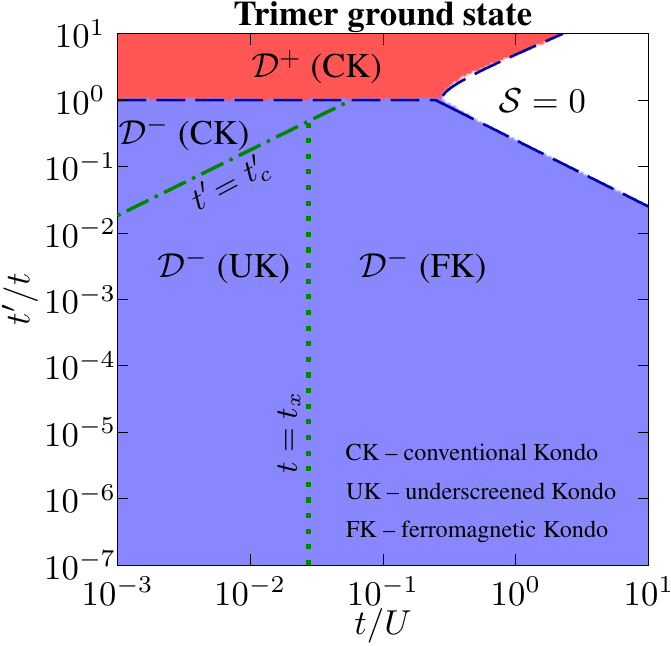}
	\caption{
		    The ground state of the isolated trimer for $\delta_i=0$ and different $t$ and $t'$.
		    The violet (red) area corresponds to the odd-parity (even-parity) doublet
		    ground state $\mathcal{D}^-$ $(\mathcal{D}^+)$,  while the white area
		    corresponds to the singlet ground state $\Spin=0$. 
		    In brackets the expected phases that emerge in the presence of coupling to the leads are
		    indicated, with schematic borders between them marked by dotted and dot-dashed lines.
		    Details are explained in \Sec{TKs}.}
	\label{fig:2a}
\end{figure}

We begin by considering the trimer decoupled from the leads. 
In general, we focus on regimes where the local Coulomb repulsion $U$ is the largest energy scale. 
The phase diagram for such case is shown in \fig{2a}.
Even though $H_{3{\rm QD}}$ can be in principle exactly diagonalized for $\delta_i=0$, the solution 
involves roots of a general quartic polynomial and is not very insightful. 
However, we find it important to note that for $\delta_i=t'=0$ the trimer Hamiltonian $H_{\rm 3QD}$,
\eq{H3QD}, exhibits the particle-hole symmetry defined by the simultaneous
transformation on all the QDi's ($i=1,\,2,\,3$),
$d_{i\s} \mapsto s_i \dk_{i\s}$, provided the coefficients $s_i$ are all of
module $1$ and $s_2=s_3=-s_1$. However, even for $\delta_i=0$, the term proportional to $t'\equiv t_{23}$ 
changes sign upon this transformation and inevitably destroys this symmetry. 
Therefore, despite the assumed local PHS at each site of the trimer, 
the global particle-hoe symmetry is not preserved,
and the trimer may not even be half-filled in the ground state. 

Insight into the energy spectrum of the trimer can be obtained based on the observation 
that, as long as $\Gamma=0$, one can use $U^{-1}$ as a small expansion parameter. One immediately sees that
there are only $8$ states of energy of the order of $-3U/2$, which are separated from the other states by energy 
differences of at least $\sim U/2$. Therefore, these are the states important for the low-temperature physics in all 
the Kondo regimes. Actually, $4$ of them form a symmetry-preserved spin quadruplet of energy 
$E_{\Spin=3/2} = -3U/2 + \delta_1 + \delta_2 +\delta_3$. The remaining states form two $\Spin=1/2$ doublets
(which are coupled to other doublets of energy at least $\sim U$ higher).
The two low-energy eigenstate doublets, denoted by $\dub{+}$ and $\dub{-}$,
are actually always lower in energy than the quadruplet and have even
($\dub{+}$) or odd ($\dub{-}$) parity with respect to exchange of QD2 with
QD3, respectively.
For $t'=0$ the ground state is the odd-parity doublet $\dub{-}$. Increasing the frustration
brought about by $t'$ causes a level-crossing QPT at
$t'=t$, as illustrated by the colored regions in \fig{2a}
(note the logarithmic scales on both axes and the $t'/t$ normalization on the vertical axis). 
For low values of $t<U/10$ and $t'<t/20$,
the energy difference $E^*=|E_{S=1/2}^{+}-E_{S=1/2}^{-}|$ of these two doublets is of the order 
of the exchange coupling between the relevant quantum dots,  
\beq
E^* \approx \frac{4t^2}{U} |1-t'/t|^2.
\label{Eex}
\eeq  
Thus, one should expect these two phases to become indistinguishable for temperatures $T \gtrsim E^*$.

Finally, as can be seen in \fig{2a}, when the inter-dot hopping becomes large in comparison to local 
Coulomb repulsion, $t,t' \gtrsim U/4$, an additional phase is present, labeled "$\Spin=0$". This is a spinless
state, occupied (for positive $t'$) with $4$ electrons. Its presence is a clear manifestation of the lack of 
the global particle-hole symmetry in the model (even in the presence of the local one),
which is caused by the frustrating coupling $t'$.
Nevertheless, this singlet is present even without coupling to the leads. It
is, therefore, not a Kondo state and will not be discussed in detail in the present paper.

\subsection{The Kondo scales}
\label{sec:TKs}

When the trimer is coupled to the leads, the most important observation concerns the effective exchange coupling
of the two doublets relevant at the lowest temperatures \cite{Mitchell_QPT}. The even doublet, $\dub{+}$,  is
coupled in a conventional anti-ferromagnetic manner, with the same strength as the QD1 spin itself,
$J^{\rm CK} = 8\Gamma/(\pi\rho U)$,
($\rho$ denotes the normalized density of leads states at the Fermi level). This means that no matter 
how weak this coupling is, whenever temperature drops below the Kondo temperature $T_K$, the trimer spin 
$\Spin=1/2$ is fully screened by the electrodes due to the conventional Kondo (CK) effect. 
Hence the CK label in \fig{2a}. The relevant value of $T_K$ can be estimated on the basis of the Anderson's
poor man's scaling method \cite{poorMan}, 
even for finite leads spin polarization $p$ \cite{Haldane,MartinekTK}, to give
\beq
T_K = 	\sqrt{\frac{\Gamma U}{2}} 
		\exp \left[ -\frac{\pi}{8} \frac{U}{ \Gamma } \frac{{\rm atanh} (p)}{p}\right] .
\label{TKp}
\eeq
For $\Gamma=U/10$ used throughout the paper one gets \mbox{$T_K(p\!=\! 0.5) \approx 0.003U$} 
and $T_K(p\!=\!0) \approx 0.0044U$.

Meanwhile, the odd-parity doublet $\dub{-}$ is coupled ferromagnetically, although with reduced strength. Namely,
$J^{\rm FK} = -J^{\rm CK}/3$
\cite{Mitchell_QPT}. Therefore, the ferromagnetic Kondo (FK) effect is expected, see \fig{2a}, which leads to asymptotically 
free spin \cite{poorMan} and singular dynamics \cite{singularFK} at low temperatures. Due to the fact that 
the exchange coupling is inevitably proportional to $\Gamma$, this gives the characteristic temperature scale
$\widetilde{T}_K$ following \eq{TKp} with $\Gamma$ replaced by $\Gamma/3$. For $\Gamma=U/10$, 
this gives $\widetilde{T}_K(p\!=\! 0.5) = 3.09 \times 10^{-7}U$ and $\widetilde{T}_K(p\!=\! 0) = 9.87 \times 10^{-7}U$.
Note, that for both two cases of $p=0$ and $p=0.5$, $\widetilde{T}_K \ll T_K$.
Therefore, one expects that in the temperature regime  of
$\widetilde{T}_K < T < T_K$ the Kondo effect takes place at QD1 only, despite quite strong $t$.
This is confirmed by NRG calculations  presented in \Sec{NRG}.

Since the ground states corresponding to CK and FK regimes differ in spin quantum number, they are separated 
by the QPT. Nevertheless, it does not occur exactly for $t'=t$. In fact, 
since $J^{\rm CK}$ scales up and $J^{\rm FK}$ scales down with decreasing temperature, it is hardly surprising
that the CK phase takes over the FK phase for $t'=t$ and the QPT line moves to $t'\equiv t'_c < t$,
where $t'_c$ denotes the transition point between the CK and FM phases, yet roughly independent of $t$.
Nevertheless, even for couplings as strong as $\Gamma=U/10$ the difference between $t'_c$
and $t$ occurs to be hardly noticeable, cf. NRG results in \Sec{NRG}. 

\begin{figure}[t]
	\centering
	\includegraphics[width=0.9\linewidth]{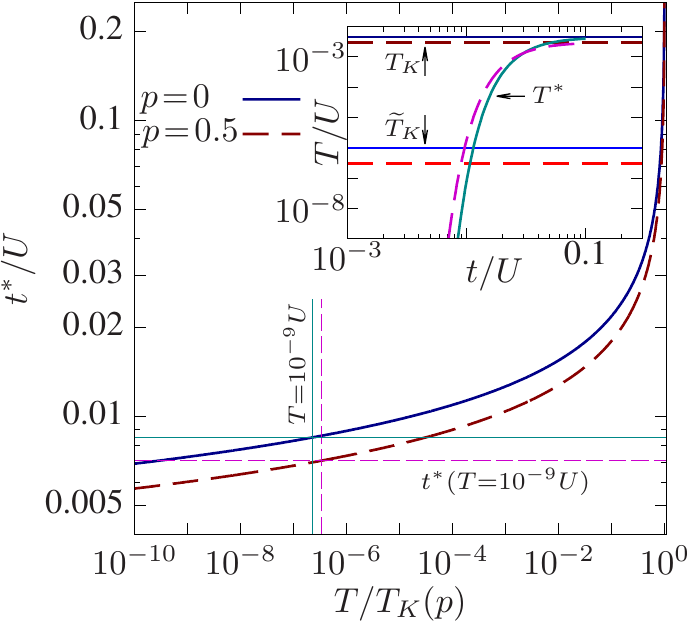}
	\caption{
		The dependence of the critical hopping $t^*$ on temperature based on \eq{tstar}
		calculated for $\Gamma=0.1U$
		in the case of non-magnetic ($p=0$) and magnetic leads ($p=0.5$).
		In the inset all the relevant Kondo scales are plotted against the hopping $t$.
		More details are provided in the main text of \Sec{TKs}.
	}
	\label{fig:2b}
\end{figure}

However, the above considerations contain an implicit assumption that the molecular trimer orbitals are still 
well-defined for $\Gamma >0$. This seems reasonable if the inter-QD exchange interactions are large in comparison 
to $T_K$, $J_2 \approx 4 t^2 / U \gtrsim T_K$. If, on the contrary, $t \lesssim \sqrt{U T_K}/2$, then at temperatures 
below $T_K$, yet above some critical value of the order of $J_2$, single electrons occupying QD2 and QD3 are 
not correlated with QD1 due to thermal fluctuations, while QD1 spin is almost fully screened by CK effect, 
therefore, forming a Fermi liquid state \cite{bookHewson}. The characteristic value of $t$, around which the 
crossover happens, shall be denoted as
\beq
t_x = \frac{1}{2} \sqrt{T_K U}.
\label{tx}
\eeq
The second and third dot (QD2 and QD3) spins may still be correlated with each other though,
if the hopping-induced anti-ferromagnetic exchange interaction ${J_2}' \sim 4 (t')^2/U$ exceeds temperature 
fluctuations. When the temperature falls 
further, also a super-exchange comes into play, mediated by QD1-and-leads quasi-free pseudoparticles.
The latter has a ferromagnetic sign and a magnitude of the order of $J_{\rm SX} \sim t^2/\sqrt{T_K U}$ 
\cite{Mitchell_phase_diag}. The interplay between $J_{\rm SX}$ and $J'_2$ (which is in fact a competition 
between $t$ and $t'$ again) determines the state of the QD2-QD3 cluster to be either the spin singlet, 
depicted in \fig{system}(b), or $\Spin=1$ triplet, cf. \fig{system}(c). 
The position of the transition anticipated from $J_{2}'=J_{\rm SX}$ criterion 
is marked in \fig{2a} with a dot-dashed line.

The story of the former case is already finished, as this is a stable low-temperature state, actually 
a special case of the CK state discussed so far. 
Yet, the fate of the triplet state is still not concluded. In fact, in the case of formation of $\Spin=1$ 
within the QD2-QD3 cluster, lowering the temperature further gives rise to another Kondo screening. 
Indeed, the QD1-and-leads Fermi liquid screens the QD2-QD3 spin at temperatures of the order of \cite{Cornaglia}
\beq
T^*(t) = \alpha T_K \exp (-\beta T_K U/4t^2),
\label{Tstar}
\eeq
as the local density of states of QD1, exhibiting the Kondo peak of the width $\sim T_K$, serves as 
a band for QD2-QD3 cluster. The coefficients $\alpha$ and $\beta$ are of the order of unity and depend 
on the system parameters weakly, see also Ref.~\cite{Cornaglia}. The dependence of $T^*(t)$ 
for $\Gamma=U/10$ and $p\in \{0, 0.5\}$ is plotted in the inset in \fig{2b}; the ($t$-independent) values of $T_K$
and $\widetilde{T}_K$ are indicated there as well. However, the screening of $\Spin=1$
cannot be complete with only one screening channel, therefore it is under-screened in the sense of 
Nozieres-Blandin Fermi-liquid theory \cite{NozieresBlandin}. Hence, we call this regime the under-screened 
Kondo (UK) regime, see \fig{2a}. It seems noteworthy, that this phase has all the quantum numbers identical to the FK phase, 
discussed  earlier, including the residual $\Spin=1/2$ spin in the ground state. In fact, 
these phases are continuously connected both for $p=0$ \cite{Mitchell_phase_diag} and $p>0$; see \Sec{NRG}. 
The estimation of the position of the UK/FK crossover based on $4t^2/U = T_K$ criterion
for $\Gamma=U/10$ and $p=0.5$ is indicated in \fig{2a} with a dotted line.

Importantly, $T^*$ given by \eq{Tstar} is very low for weak $t'$, so that at some temperature $T>0$, 
there exists such a critical value of $t$, denoted $t^*(T)$, that $T^*[t^*(T)]=T$. In fact, taking 
$\alpha\approx\beta\approx 1$ we find from \eq{Tstar}
\beq
t^*(T) \approx \frac{1}{2} \sqrt{ \frac{T_K U}{ \log( T_K / T ) }}
\label{tstar}
\eeq
Estimating the Kondo temperature from \eq{TKp}, for $\Gamma=0.1$ and $p=0.5$, one can calculate $t^*$ for 
experimentally relevant temperatures and make clear that in practice for $t^* < 0.005U$ 
the non-zero temperature regime is experimentally relevant, see \fig{2b}.

As explained earlier, the transition point between the CK and FK phases, $t'_c$, remains practically independent of $t$ 
and close to $t'=t$. However, this is no longer the case in the UK regime, where the transition is strongly shifted 
to \cite{Mitchell_phase_diag}
\beq
t_c' \approx\frac{ t^2 }{\sqrt{T_K U}},
\label{tc}
\eeq 
which is particularly small for weak $t$. 
This estimation of transition point is indicated in \fig{2a} by the dot-dashed line.
Note that due to the dependence of the Kondo temperature on spin polarization
$p$ the critical value $t_c$ is a function of $p$ as well.
Furthermore, due to the fact that the UK and FK phases are separated by a continuous crossover only, 
it is sensible to continue the line to the transition position characteristic of the FK regime.

\subsection{The exchange field}
\label{sec:ex}

In general, the coupling between a nano-device 
and the leads gives rise to the renormalization of 
the energy levels of the nano-device. In the case of magnetic leads, 
this renormalization is usually spin-dependent 
\cite{MartinekEx}. 
The part of its linear contribution proportional to 
leads magnetization $p$ is often called the (spintronic) 
exchange field and will be denoted $\ex$ \cite{KW-ex}. 
For single impurity $\ex$ is altered smoothly while 
lifting impurity energy level, changing sign at the 
local PHS point. However, in the trimer case PHS 
is broken by the frustration, and $\ex$ no longer 
vanishes even at local PHS, opening the possibility for 
spin polarization of the nanostructure in such conditions.
It is noteworthy that at sufficiently small temperatures 
even very small value of frustrating hopping $t'$ may result in substantial magnitude of $\ex$.
In the present section to obtain certain insight into the properties of $\ex$
in the case of locally PHS trimer we performed a perturbative  analysis.
However, these predictions will be corroborated 
further in \Sec{NRG} by accurate NRG calculations.

For each eigenstate of the isolated trimer $\ket{e_i}$, the shift of its energy $E_i$ due to interaction with 
ferromagnetic leads is linear in $p$ in the leading (second) order of perturbation theory in the hopping
matrix elements between the trimer and the leads. Within the wide-band limit discussed in \Sec{model}, 
the exchange field in that state is, therefore, defined as \cite{KW-ex}
\beqa
\ex^i \es  \sum_{j\s}  \frac{\s p \Gamma}{\pi \bra{e_i}\hat{\Spin}_z\ket{e_i}} \log \left| \frac{ E_j-E_i }{ D + (E_j-E_i) } \right| \times \nn\\&&
	\times \left( |\la e_j| \dk_{1\s} |e_i\ra|^2 + |\la e_j| d_{1\s} |e_i\ra|^2  \right)  ,
\label{ex}
\eeqa
where the spin index $\s$ is understood as $\pm 1$ when factoring numbers, and $\hat{\Spin}_z$ denotes
the operator of $z$th component of trimer spin. This is a proper definition for 
$\bra{e_i}\hat{\Spin}_z\ket{e_i} \neq 0$, yet for spin-less states the right-hand side of the equation vanishes
anyways and $\ex^i$ can be set arbitrarily. The convenient choice is to put it to the mean over the values
within the multiplet for $\Spin > 0$ states and to $0$ for $\Spin=\Spin_z =0$. 
In fact, the structure of the low-energy spectrum presented 
in~\Sec{3QD}, \ie spin quadruplet and two spin doublets, is preserved within this definition, namely
$\ex^{i}$ is the same for all states within each multiplet (but differs between multiplets).

\begin{figure}[t]
	\centering
	\includegraphics[width=1\linewidth]{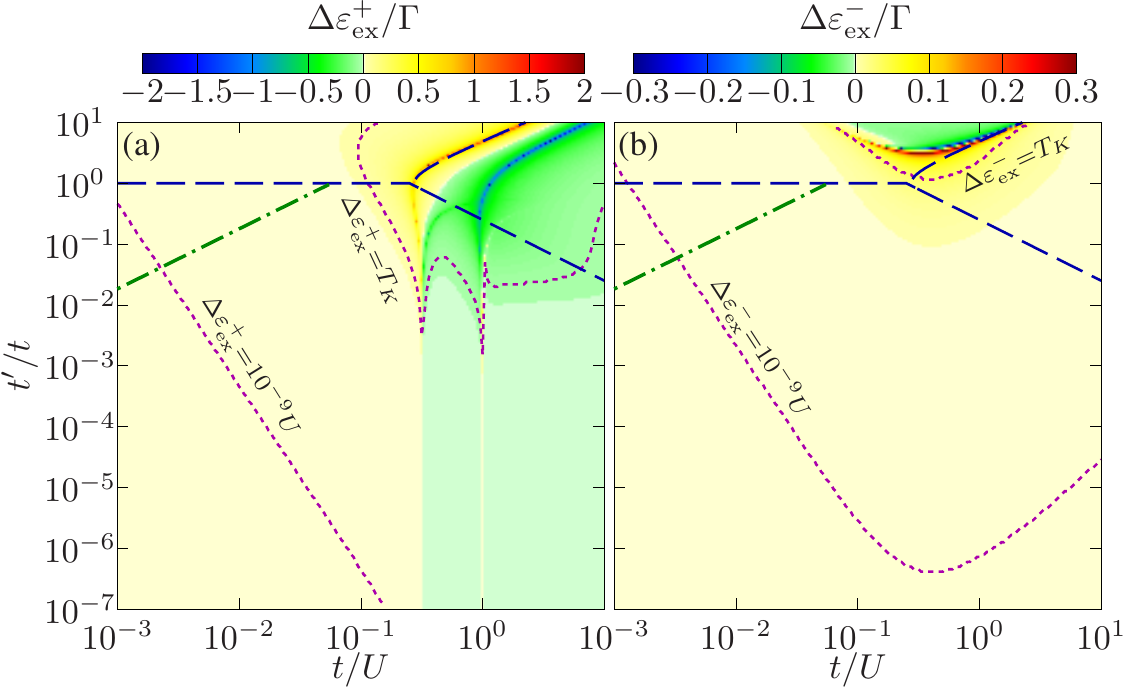}
	\caption{
		(a) The exchange field in the even-parity doublet state $\dub{+}$, calculated within perturbation 
		theory for $U=D$ and $p=0.5$. 
		(b) Similar plot of the exchange field in the odd-parity doublet state $\dub{-}$.
		Note the logarithmic scales on both axes in all plots. 
		More details are provided in the main text of \Sec{ex}.
	}
	\label{fig:ex}
\end{figure}

The exchange fields in the two relevant doublet states, $\dub{+}$ and $\dub{-}$,
denoted correspondingly $\ex^+$ and $\ex^-$, 
are presented in Figs.~\ref{fig:ex}(a)-(b) for a trimer at local PHS point in a wide range of $t$ and $t'$.
This wide range of hopping constants allows for making predictions for different possible realization of the trimer, 
including molecules as well as quantum dot systems. Note in particular, that the smallest used value of $t'$, $t'/t=10^{-7}$, is
already a value one can expect for a next-nearest-neighbor interaction strength in a linear molecule. 

The dashed and dot-dashed lines shown in Figs.~\ref{fig:ex}(a)-(b) are the same as those in \fig{2a}
and indicate the positions of the QPTs.
The first observation is that in the regimes where the scales comparison suggest
the CK ground state, \ie where $\dub{+}$ is the most relevant state, $\ex^+ > 0$. Similarly, wherever the UK or FK ground
state is expected, $\ex^- > 0$, while the exchange field in the spin-less ground state obviously vanishes, 
$\ex^{\Spin=0}=0$.
Therefore, one expects that at local PHS (assumed for the calculation) the exchange field in the states relevant 
at low $T$ is non-negative, $\ex \geq 0$. This is in agreement with transport properties calculated with NRG
in \Sec{NRG}. 
Moreover, one also expects that whenever the ground state of the system is not a spin singlet,
$\ex$ would split the ground state spin degeneracy and lead to states comprising a spin polarized
trimer. This should be expected in particular in the FK and UK phases,
which from now on will be denoted 
as FK$'$ and UK$'$, when the doublet degeneracy is lifted.
Furthermore, $\ex > T_K$ may split the Kondo resonance in the CK phase \cite{MartinekTK,MartinekEx}.
In the following this resulting regime will be denoted as CK$'$. 
In other words, the different regimes in the case of magnetic leads
will be denoted with a prime.

One easily notes that $\ex$ is in general quite small, except for the regions where two relevant states are close to
degeneracy, since then the denominator in \eq{ex} blows up. However, as this is only a perturbative expression, 
one should take that result with a lot of caution. Even though some enhancement of trimer energy levels 
renormalization is expected there, one does not, in general, expect them to be divergent, even in the $T\to 0$ limit. 
Indeed, note that NRG results presented in \Sec{NRG} indicate regular behavior of the trimer magnetization.

Finally, from Figs.~\ref{fig:ex}(a)-(b) it is evident that the exchange field in the ground state, $\ex^{\rm GS}$, 
has apparently quite a small absolute value. To make it even more clear, we added dotted lines in both figures
to indicate where the exchange field is equal to the conventional Kondo scale, $\ex^{\rm GS} = T_K$,
and where it equals $10^{-9}U$. The latter is intended to mimic the zero-temperature regime. Clearly, even at such 
a small $T$ not for all of the considered parameters $|\ex^{\rm GS}| > T$ is expected. This feature occurs
important for the phases of trimer in all temperature regimes. 
In particular, one can expect that the phases FK$'$ and UK$'$
are possible only when $T<\ex$, while at elevated temperatures the non-magnetized 
FK and UK regimes should be expected.

\section{The overview of the phase diagram}
\label{sec:PD}

\begin{figure}[tb!]
\centering
\includegraphics[width=1\linewidth]{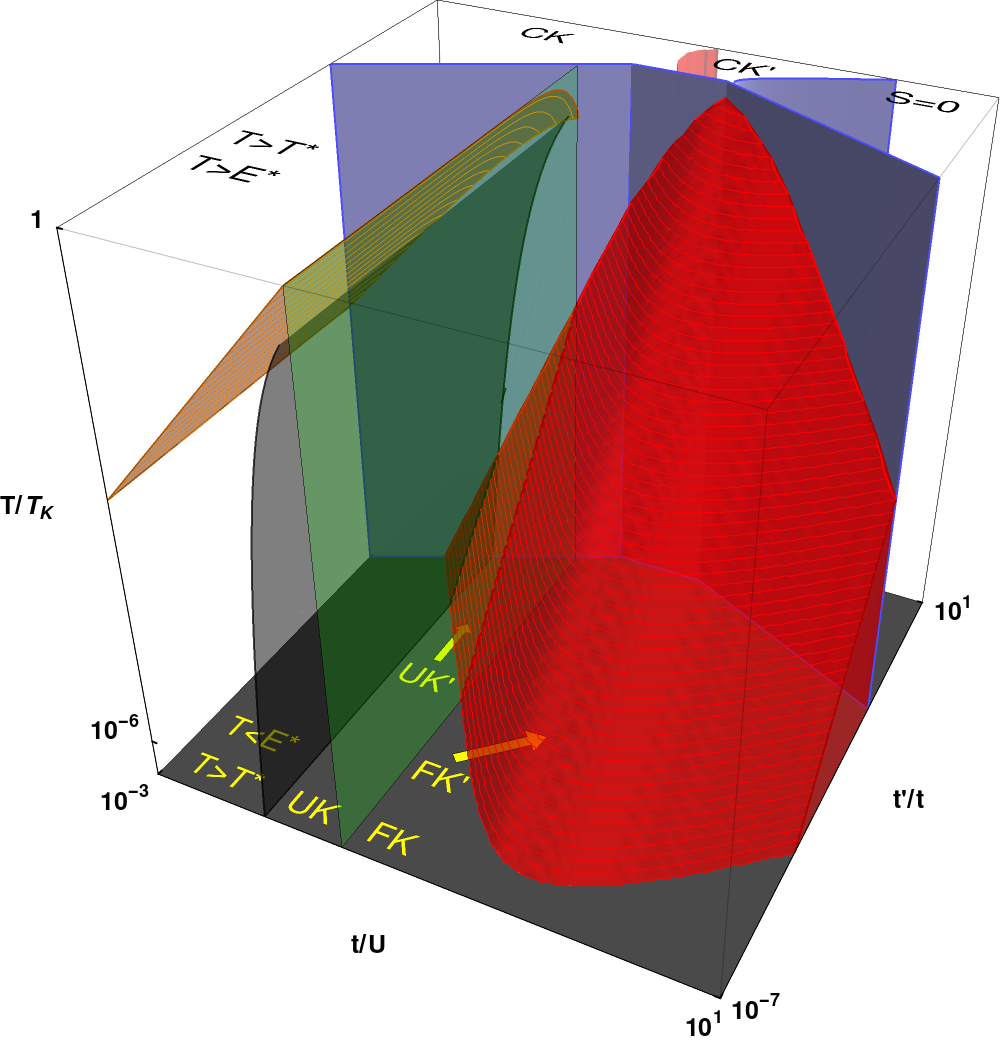}
\caption{
		 Schematic phase diagram of the considered system for $T\leq T_K$ and for
		 magnetic leads with spin polarization $p=0.5$
		 calculated vs the following parameters:
		 the hopping $t$ (in units of Coulomb repulsion $U$), 
		 the frustrating coupling $t'$ (normalized by $t$), 
		 and the temperature $T$ (scaled by $T_K$).
		 Note the logarithmic scale of all the axes.
		 The respective phases are described in the main text of \Sec{PD}.
		 }
\label{fig:PD}
\end{figure}

The $3$-dimensional phase diagram of the trimer, featuring $t$, $t'$ and $T$
as parameters, is presented in \fig{PD}. Already the first sight of it allows to realize that it is fairly 
complicated, however, the analysis of energy scales performed
in the preceding section shall allow us to identify 
and understand all the phases. 

We start the discussion from the QPT lines introduced 
as dashed or dot-dashed lines in \fig{2a}. They are presented as solid vertical walls, without any broadening 
for elevated temperatures for the sake of clarity of the figure. Their positions are based on the exact positions of
ground state changes for the isolated trimer for $t>\sqrt{T_K U}$ and given by \eq{tc} for smaller $t$.
The transparent vertical wall is used to indicate the position of the crossover between UK and FK phases,
which is quite arbitrarily defined to be at $t=t_x$ fulfilling \eq{tx}.

In turn we move to the discussion of non-zero $T$ properties of the UK phase. As explained in \Sec{TKs}, the
second Kondo temperature is indeed cryogenic for small $t$. The approximate position of the crossover between
partially screened and unscreened $\Spin=1$ QD2-QD3 cluster is indicated in \fig{PD} by the dark opaque leaning 
surface, based on \eq{tstar}. Even though the bottom of the figure corresponds to $T=10^{-9}U < 10^{-6}T_K$,
the uttermost left part of the figure still corresponds to the $T>T^*$ regime, which vanishes only in the purely
mathematical $T\to 0$ limit. On the other hand, further increase of $T$ inevitably leads to the next crossover,
occurring when the thermal energy reaches the excitation energy between the two relevant eigenstate doublets,
$T = E^*$. Above this threshold, the states at two sides of the transition are similarly probable and the
physical properties are expected to be a mean of the properties of each of them. In particular, the $\Spin=1$
state is not fully formed within the QD2-QD3 cluster. Additionally, 
note that $E^*$ is estimated by \eq{Eex}, however, one needs to take into account that this formula does not
take into account the shift of the UK/CK quantum phase transition away from $t'=t$, therefore it overestimates
$E^*$ very close to that transition. Nevertheless, this estimation is sufficient for qualitative understanding 
of the phases of the system and is used to plot the crossover position as a skewed surface in the phase diagram
in \fig{PD}. 

It is noteworthy to point out that all the phases discussed so far exist also for $p=0$.
However, some changes in position of borders occur then, because of
the difference between (lower) $T_K(p\!=\!0.5)$ and (higher) $T_K(p\!=\!0)$, cf.~\eq{TKp}. Therefore, for example,
$t'_c$ is somewhat smaller for $p=0$, as follows from \eq{tc}. Similarly, $t^*$ is larger for $p=0$, 
cf.~\fig{2b}.

Another way to obtain interesting spintronic properties is to exploit the unique features present only 
for $p>0$. They are in general caused by the presence of the exchange field in the ground state, 
$\ex^{\rm GS} \neq 0$. First of all, the exchange field suppresses the CK effect if $\ex \gg T_K$ and splits the Kondo 
peak in QD1 spectral density for $\ex \approx T_K$. In both cases one expects QD1 to become spin polarized,
even though at local PHS (as considered here) the global PHS is broken actually only by the $t'$ 
hopping between QD2 and QD3. Therefore, one can see the coupling to the QD2-QD3 cluster as a kind of functionalization
of QD1-based device. These magnetic phases are separated from basically non-magnetic state for 
$\ex^{\rm GS} \ll T_K$ by a continuous crossover, as it is for the case of a single quantum dot outside
of the PHS point \cite{MartinekEx,MartinekTK}, indicated in \fig{PD} with a curved vertical wall, with magnetic 
phase labeled as CK$'$ and the non-magnetic simply by CK on the top face of the diagram.

Furthermore, one can predict even more pronounced effect of the exchange field at the FK side of the transition.
There, not only is the relevant Kondo scale much smaller, but also the ground state comprises asymptotically 
free spin doublet, so at sufficiently low temperatures the exchange field always overcomes the ferromagnetic 
coupling to the leads. Therefore, in the $T\to 0$ limit only the phase with non-zero dots magnetization, 
denoted FK$'$, is stable. However, as discussed in \Sec{ex}, the magnitude of the exchange field is actually 
very small for small $t$ and $t'$, so that at finite temperatures the region where the thermal fluctuations do not 
overcome $\ex^{\rm GS}$ is finite, compare dashed lines in Figs.~\ref{fig:2a} and \ref{fig:ex}.
This gives rise to the crossover between the magnetic phase FK$'$
and the non-magnetic FK phase at $T = \ex^{\rm GS}$, which is
indicated in \fig{PD} with a striped dome-like surface. Note also, that dotted lines labeled as 
"$\ex^- = 10^{-9}U$" in \fig{ex}(b) signify in fact the footprint of the FK$'$ phase on the "floor" of the diagram,
corresponding to $T=10^{-9}U$.

Both FK and FK$'$ phases continue through the described earlier crossover toward the UK (and correspondingly the magnetic 
UK$'$) phase, where additionally effective $\Spin=1$ state is formed within the QD2-QD3 cluster. This is particularly 
interesting state, as here QD1 in fact experiences CK, yet still in partially-screened QD2-QD3, the magnetic order 
is imposed, with $\Spin=1$ almost fully aligned with leads minority spins for temperatures both below $T^*$ and above it. 
This is the case as long as the temperature does not overcome the effective ferromagnetic coupling between 
the second and third quantum dot.

\section{Numerical results}
\label{sec:NRG}

In this section we present the results of NRG calculations concerning the physical properties representative
for each Kondo regime of the system. These include the linear conductance $G$ and the expectation value of the 
trimer spin $\Spin$, the trimer's spin polarization, as well as the trimer's entropy.
The studied quantities clearly confirm 
the predictions of the qualitative analysis performed in Secs.~\ref{sec:E} and \ref{sec:PD}.

\subsection{Conductance}
\label{sec:G}

\begin{figure}[t]
\centering
\includegraphics[width=1\linewidth]{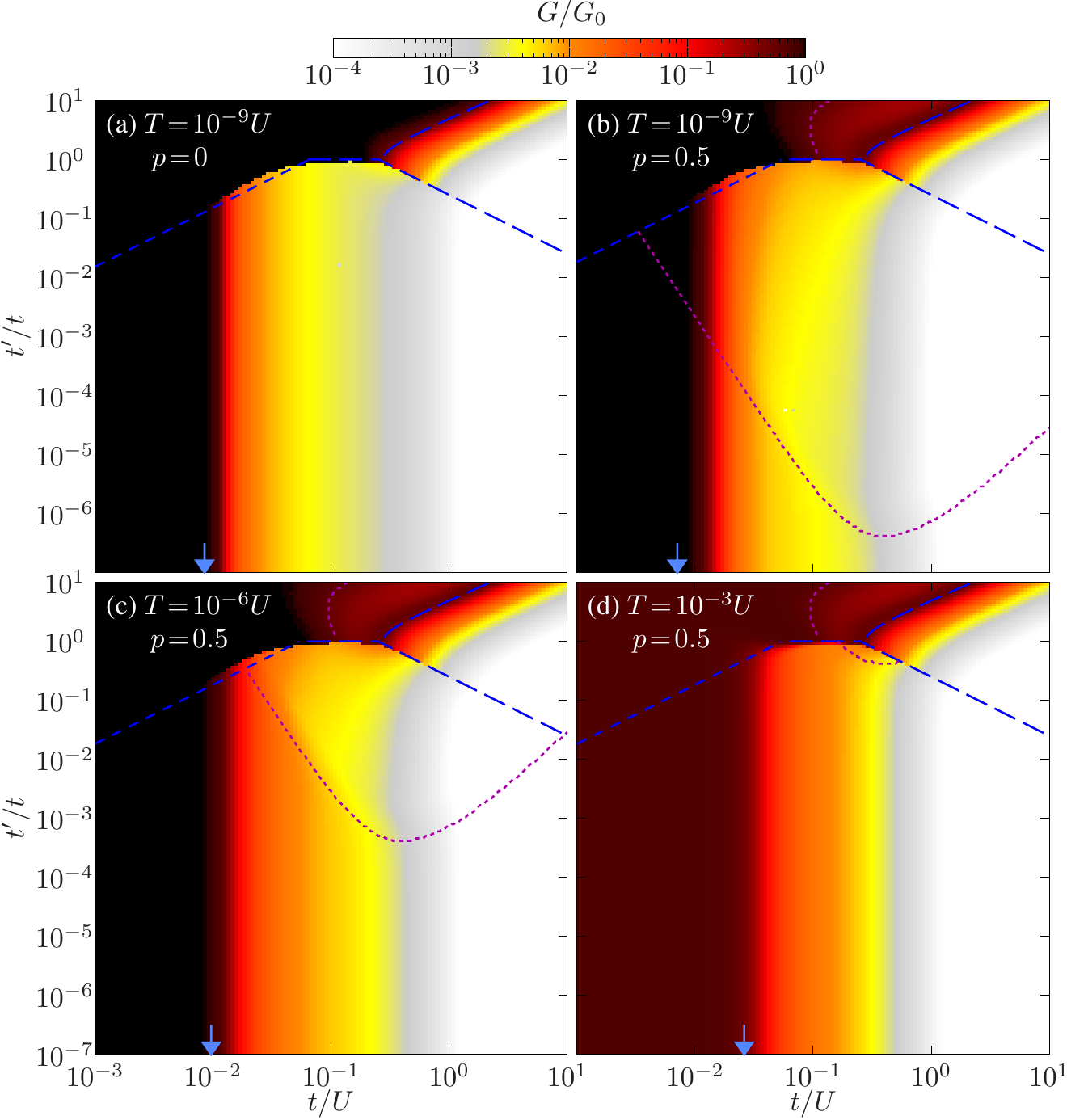}
\caption{
		 Conductance as a function of $t$ and $t'$ for $\Gamma=0.1U$, $\delta_i = 0$ 
		 and for (a) $T=10^{-9}U$ and $p=0$
		 (b-d) finite spin polarization $p=0.5$ and different temperatures
		 indicated in the figure.
		 Dashed lines correspond to boundaries of phases from \fig{2a} and \fig{PD}.
		 Arrows indicate $t=t^*$ points on vertical axes according to \eq{tstar}.
		 Note the logarithmic color-scale.
		 }
\label{fig:G2D}
\end{figure}

The unitary conductance through the device is possibly the most well-known hallmark of the conventional Kondo state
for nonmagnetic leads, $G=G_0 = 2e^2/h$. The conductance possesses this value in the CK regime and in the $T>T^*$ part 
of the UK regime, see \fig{G2D}(a). On the contrary, it abruptly changes at the QPT between
the CK and FK phases, while changing continuously with increasing $t$ from the UK to the FK phase.
In agreement with earlier predictions, 
at the CK side of the transition the conductance remains maximal while increasing $t$ up to the transition point to the 
$\Spin=0$ phase, where it ultimately vanishes. Notably, for $p=0$, there is hardly any $t'$ dependence 
of the conductance for $t'<t'_c$.

The latter is no longer true for $p>0$. It is clearly visible in \fig{G2D}(b), that in the FM$'$ region
[whose border is indicated with a dashed line, similarly to Figs.~\ref{fig:ex}(a)-(b)] the conductance depends on $t'$, 
and is in particular larger than for $p=0$. This trend persists also at higher $T$, cf. \fig{G2D}(c). 
However, for sufficiently large temperatures, the FM$'$ region is practically not present; see \fig{G2D}(d). 
Meanwhile in the CK regime for $p=0.5$,
$G = G_0$ and is reduced after crossover to the CK$'$ phase driven by increasing $t$. This remains true as long as 
$T\ll T_K$, as visible in Figs.~\ref{fig:G2D}(b)-(c). However, since $T=10^{-3}U$ is already close to $T_K$, the conductance 
in the CK phase drops in this case below $G_0$ and decreases even further in the CK$'$ regime.

\subsection{Spin expectation value}
\label{sec:Spin}

\begin{figure}[t]
\centering
\includegraphics[width=1\linewidth]{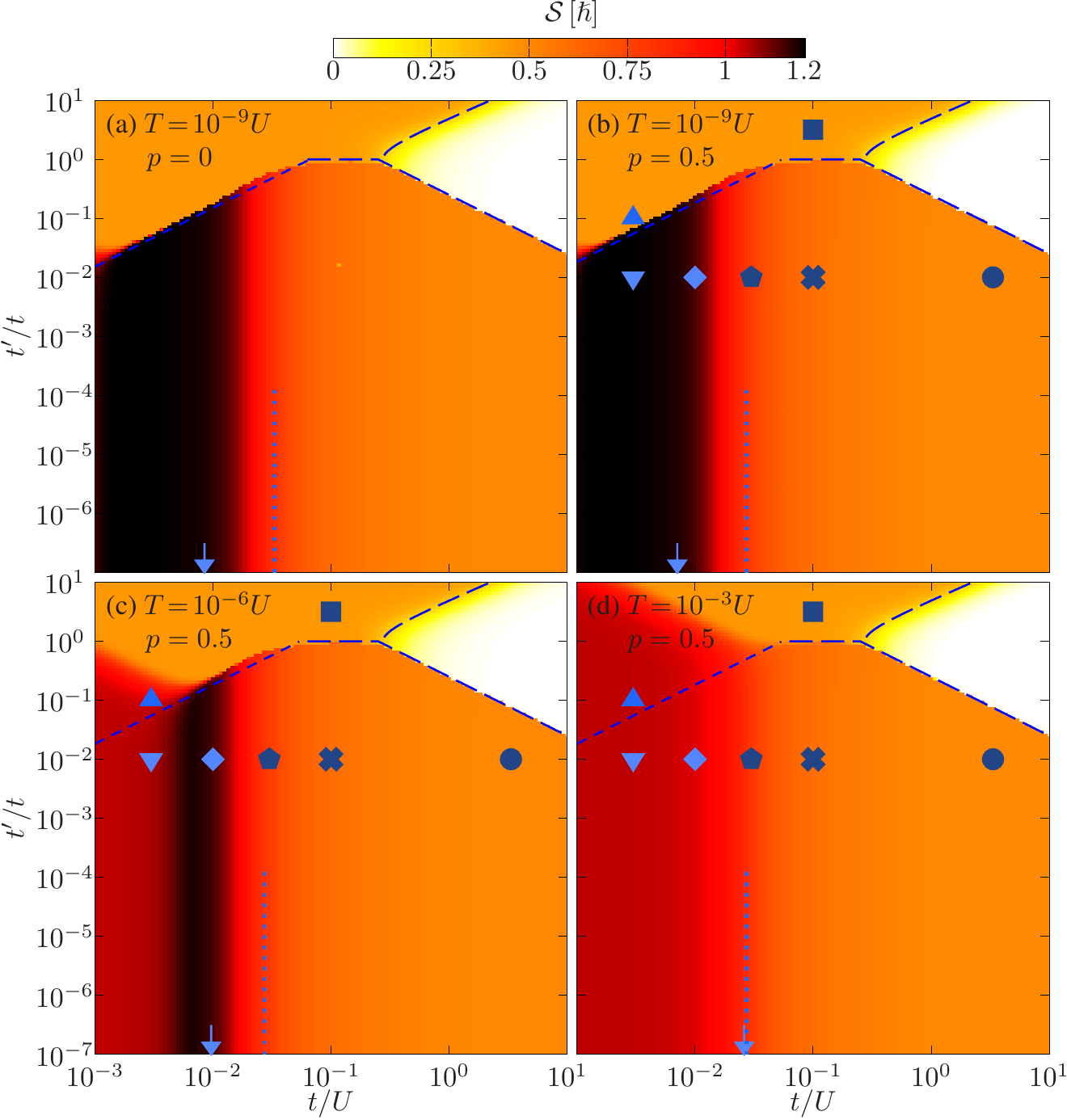}
\caption{
		 The expectation value of the trimer spin $\Spin$ as a function of $t$ and $t'$.
		 The parameters are the same as in \fig{G2D}.
		 The points marked with symbols indicate $t$ and $t'$ for which \fig{S1D} is prepared.
		 Small vertical arrows (dotted lines) indicate $t^*$ ($t_x$) positions, correspondingly. 
		 }		 
\label{fig:S2D}
\end{figure}

\begin{figure}[t]
\centering
\includegraphics[width=0.9\linewidth]{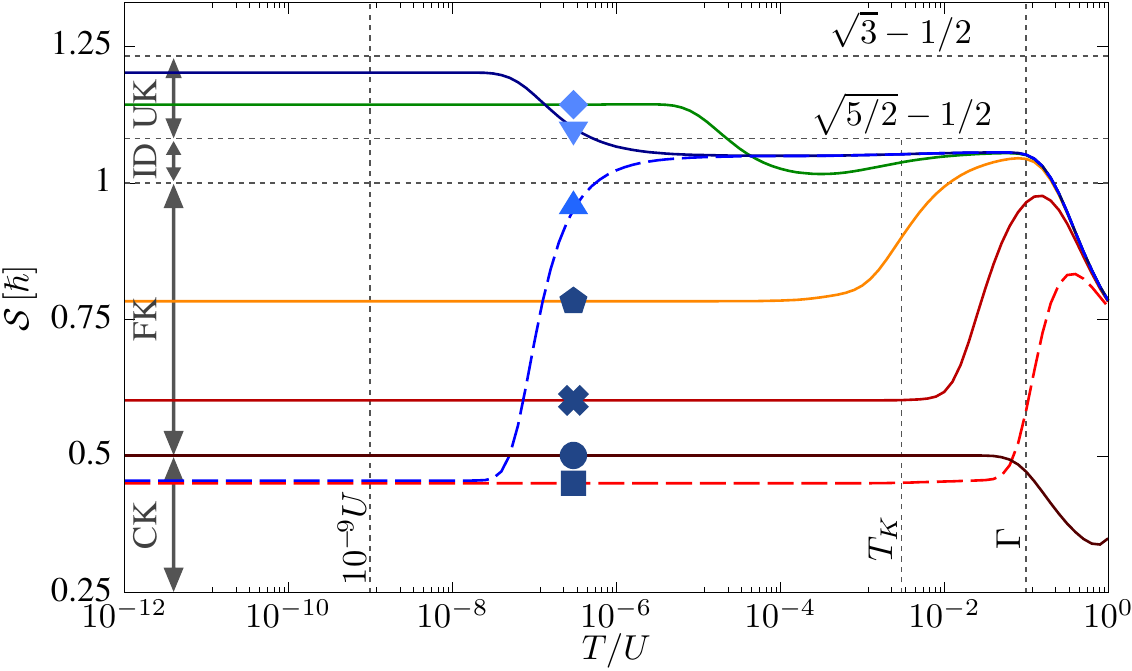}
\caption{
		 The trimer spin expectation value $\Spin$ as a function of $T$ for values of $t$ and $t'$ indicated in
		 Figs.~\ref{fig:S2D}(b)-(d) with corresponding symbols. The other parameters are the same as in \fig{S2D}. 
		 See \Sec{Spin} for details.
		 }		 
\label{fig:S1D}
\end{figure}

In the behavior of the conductance it is not possible to see the difference between the CK and UK phases
in the regime of $t<t^*$. 
Therefore, we now analyze the expectation value of the trimer spin, defined as such a scalar $\Spin$ that 
the expectation value of the operator of trimer squared spin, $\hat\Spin_{3 {\rm QD}}^2$, fulfills
\beq
\Spin (\Spin+1) = \mean{\hat \Spin_{3 {\rm QD}}^2} .
\label{Sdef}
\eeq
This definition allows us to talk about \emph{trimer spin} as a continuous quantity, in principle having values 
in the range $\{ 0 \leq \Spin \leq 3/2\}$. Note, that Kondo screening of the local moment does not lead 
to screening of the spin in terms of the definition given by \eq{Sdef}. This is because the leads states (also these 
screening local spins) are averaged out when calculating the expectation value. Therefore, 
$\Spin$ quantifies the magnitude of the spin screened in the Kondo phase, rather than the degree of screening.

\subsubsection{Conventional Kondo (CK) regime at low temperatures}

Keeping that in mind, for low temperatures one expects in particular $\Spin\approx 1/2$ in the CK phase.
However, as one can see in Figs.~\ref{fig:S2D}(a) and (b), the CK phase value 
of $\Spin$ is in fact somewhat smaller than $1/2$ and close to $\Spin=0.45$.
This is because for large $t'$, the QD2-QD3 effective exchange 
has anti-ferromagnetic nature and the charge fluctuations are more likely to cause the $\Spin=0$ state to be intermediate state
[with empty QD1 and QD2-QD3 in a singlet state, cf. \fig{system}(b)] than the $\Spin=1$ state [cf. \fig{system}(c)].
This is indeed confirmed in \fig{S2D}(a) for $p=0$ and in \fig{S2D}(b) for $p=0.5$, see in particular
the points indicated by the square and the up-turned triangle in the latter. 
Apparently, except for very small changes in the positions of phase boundaries,
the degree of spin polarization $p$ is
pretty much irrelevant for the spin expectation values 
(this is obviously not true for $\Spin_z$, see \Sec{Sz}). 

\subsubsection{Conventional Kondo phase at higher temperature}

The temperature dependence of $\Spin$ for $t$ and $t'$ corresponding to these two points is presented 
in \fig{S1D} with dashed lines and adequate symbols.
One clearly sees that while for $t=0.1U$ and $t'=3t$ (square) $\Spin(T)$ remains constant up to 
$T\sim\Gamma$, while for $t=0.003U$ and $t'=0.1t$ (up-turned triangle) the spin expectation value rises already 
for $T\sim 10^{-5}T_K$. The latter is caused by the fact that internal trimer exchange couplings and the 
excitation energy $E^*$ are all very small for weak $t$ and $t'$, cf. \eq{Eex} and the discussion following 
\eq{tx}. Therefore, for $T>E^*$, the magnetic correlations between the individual quantum dots become irrelevant
and all the states comprising singly occupied dots are almost equally probable. There are $8$ such states,
forming two $\Spin=1/2$ doublets and a single $\Spin=3/2$ quadruplet, thus for $E^* \ll T \ll U$ we have
$\mean{\hat \Spin_{3 {\rm QD}}^2} 
= 9/4$
and hence the universal middle-temperature value for small values of $t$ and $t'$ is $\Spin = \sqrt{5/2}-1/2 \approx 1.08$. 
As seen in \fig{S1D}, in reality it is somewhat smaller due to the residual correlations, 
nevertheless \figs{S2D}(c)-(d) show how wide is the range of parameters where this formula holds. 
%
%
Note however, that while quantum dots are not correlated
among themselves, QD1 may still be Kondo-screened by the leads.

\subsubsection{Underscreened Kondo (UK) phase}

Clearly, the $\Spin \approx \sqrt{5/2}-1/2$ region includes also states belonging to the UK phase at temperatures 
above $E^*$, \ie when the effective $\Spin=1$ state is not yet formed in the QD2-QD3 cluster; see the lines denoted 
by a down-turned triangle and a pentagon in \fig{S1D} and their position in \figs{S2D}(b)-(d). However, at low 
temperatures $\Spin$ approaches another quite universal value, $\Spin \approx \sqrt{3}-1/2 \approx 1.23$. 
Again, the true maximum is slightly smaller, see the corresponding lines in \fig{S1D}, but the increase from 
below $\Spin\approx \sqrt{5/2}-1/2$ is clearly visible. This value can also be understood as characterizing the trimer
comprising QD2 and QD3 forming spin triplet and QD1 forming spin doublet state. Averaging over possible
$z$-th component configurations gives $\mean{\hat \Spin_{3 {\rm QD}}^2} = 11/4$, \ie $\Spin=\sqrt{3}-1/2$.
Therefore, this value (slightly decreased by remaining correlations) is characteristic of the underscreened Kondo phase.

\subsubsection{Ferromagnetic Kondo (FK) phase}

Since the UK phase is separated from the FK phase only by the crossover, the value of $\Spin$ decreases continuously 
towards $\Spin=1/2$ with increasing $t$. However, opposite to the CK phase, the residual QD2-QD3 correlations are
ferromagnetic in this regime, therefore, the final value $\Spin \gtrsim 1/2$, as can be seen in \fig{S1D} for 
the curves marked with a pentagon, cross and a circle.

\subsubsection{Summary of the section}

In summary, the spin expectation value $\Spin$ defined in \eq{Sdef} is an excellent marker of the phases, 
capable of differentiating between all the relevant regimes, especially these having similar transport properties. 
It reaches the highest values $\Spin \lesssim \sqrt{3}-1/2$ in the UK phase (both below and above $T^*$).
It is reduced below $\Spin \approx \sqrt{5/2}-1/2$ in the regime of almost independent, but singly occupied
quantum dots and decreases continuously toward $\Spin \gtrsim 1/2$ in the FK phase. The QPT
between the FK and CK regions is marked by an abrupt jump of $\Spin$ to some value $\Spin \lesssim 1/2$ on the CK side of the 
transition. Finally, the spinless trimer phase, scarcely discussed here, is characterized by $\Spin=0$.
$\Spin$ does not change due to the Kondo screening, yet its presence or absence in each phase can be recognized 
from the value of the conductance, as explained in the preceding section and visible in \fig{G2D}. 
The nature of different phases is further confirmed in \Sec{entropy} by the calculations of the trimer's entropy.
However, the spin expectation value does not allow for distinguishing the polarized phases 
from their non-polarized counterparts, \ie CK from CK$'$,
FK from FK$'$ and UK from UK$'$. Therefore, there is one additional quantity one needs in order to pinpoint all the
phases within NRG framework, namely, the trimer spin polarization. The related results are presented in \Sec{Sz}.

\subsection{Trimer spin polarization}
\label{sec:Sz}

\begin{figure}[t!]
\centering
\includegraphics[width=1\linewidth]{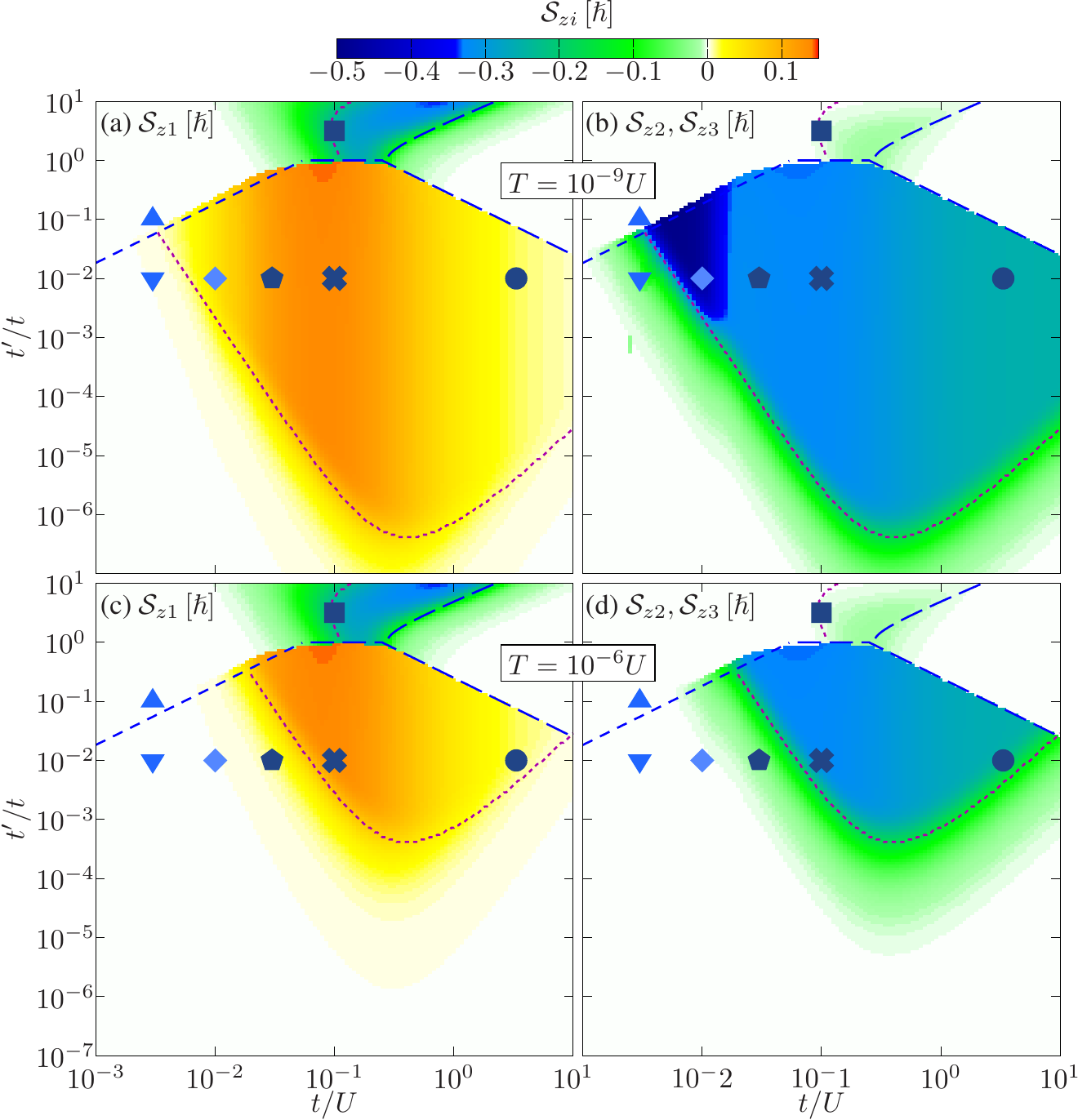}
\caption{
		 The $z$-component of spin of respective quantum dots as functions of $t$ and $t'$. 
		 Parameters are the same as in \fig{G2D} with $p=0.5$.
		 Dotted lines indicate where the condition $\ex^{\rm GS}=T$ (in the UK or FK phase)
		 or $\ex^{\rm GS} = T_K$ (in the CK regime) is fulfilled.
		 Symbols have the same positions as in \fig{S2D}.
		 }		 
\label{fig:Sz2D}
\end{figure}

\begin{figure}[t!]
\centering
\includegraphics[width=0.9\linewidth]{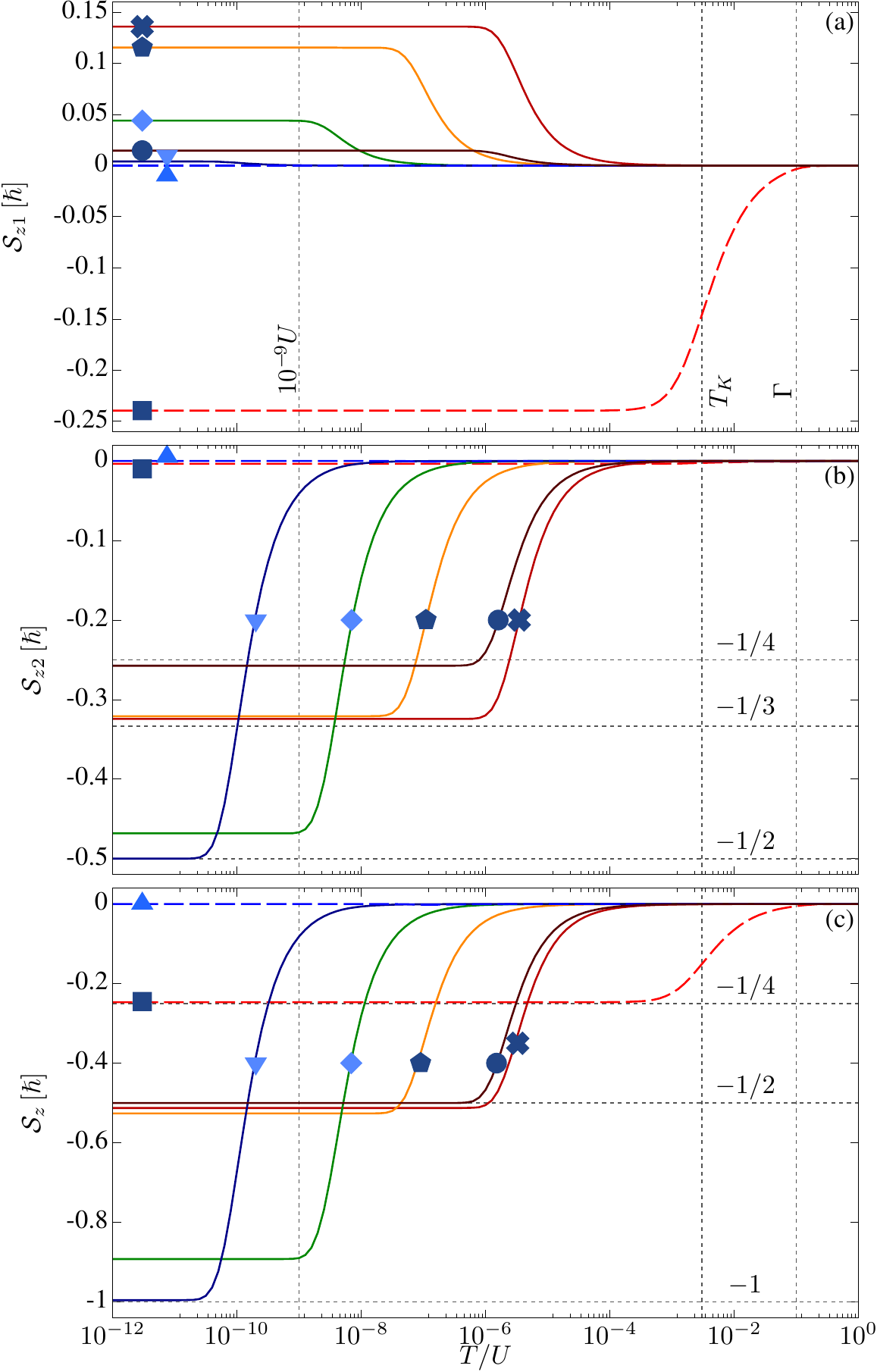}
\caption{
		 The $z$-component of spin of respective quantum dots as function of $T$ 
		 for values of $t$ and $t'$ indicated in
		 \fig{Sz2D} with corresponding symbols.
		 The other parameters are the same as in \fig{Sz2D}. 
		 See \Sec{Sz} for details.
		 }		 
\label{fig:Sz1D}
\end{figure}

An important consequence of the existence of the exchange field is the spin polarization of the trimer, 
quantified by the expectation value of the spin of respective quantum dots, denoted $\Spin_{zi}$ for QD$i$.
As can be seen in the left column of \fig{Sz2D}, $\Spin_{z1} \neq 0$ in the UK and FM phases, as long as 
the condition $T<\ex^{\rm GS}$ is fulfilled. It seems noteworthy that $|\Spin_{z1}|$ does not reach $\pm 1/2$,
yet the values are typically of the order of $1/10$, even for very small values of frustrating coupling $t'$; 
cf. \fig{Sz1D}. 
In fact, in the $T\to 0$ limit $\Spin_{z1} \neq 0$, in the whole UK and FK regimes for any non-zero $t'$. This is 
in contrast to the case of a single quantum dot slightly detuned from the particle-hole symmetry. Then, 
the quantum dot spin polarization is proportional to the symmetry-breaking detuning. It also means that 
the ground state always belongs to the spin-polarized phase (CK$'$, FK$'$ or UK$'$), unless the system is tuned
into the spinless $\Spin=0$ phase, cf. \fig{2a} and \fig{PD}.

Remarkably, in the CK and CK$'$ phases $\Spin_{z1} \leq 0$, \ie it has a tendency to align anti-parallelly 
to the leads majority spins. The QD1 spin polarization is strong in the CK$'$ phase, while it almost vanishes for the CK one.
On the contrary, in the FK/FK$'$ regime the exchange coupling to the leads changes sign, hence 
$\Spin_{z1} \geq 0$. Again, in the FK$'$ phase the absolute value of $\Spin_{z1}$ is reasonably large and does not 
vanish even for very small values of frustrating coupling $t'$, while in the FK state it is exponentially suppressed
by non-zero temperature. 

Similarly to the other regimes, in the UK$'$ phase $|\Spin_{z1}| \gg 0$, while in the UK phase $\Spin_{z1}$ almost 
vanishes. However, somewhat counter-intuitively, $\Spin_{z1} \geq 0$ also in the UK/UK$'$ phase 
(that is, the sign is opposite to the one in the CK phase), even though at elevated $T>T^*$ the conventional 
Kondo screening takes place there. This is a consequence of the fact that the sign of the exchange
field is related to detuning from the particle-hole symmetry. In the model considered in the present
paper the particle-hole symmetry is broken only by $t'$. Therefore, the formation of the exchange field 
(also at QD1) is governed by the molecular trimer states and the sign of $t'$. As noted in \Sec{ex}, 
for $t'>0$, in the ground state, $\ex^{\rm GS} > 0$, therefore, the \emph{total} trimer spin $z$-component, 
$\Spin_{z}$, is always negative. What changes between the phases is that in the CK/CK$'$ phase the trimer spin 
consists almost exclusively of QD1 spin, $\Spin_{z}\approx \Spin_{z1}$,
while in the FK/FK$'$ and UK/UK$'$ phases QD2 and QD3 form triplet instead of singlet state and 
$\Spin_{z}\approx \Spin_{z2}+\Spin_{z3}-\Spin_{z1}$. Consequently, the sign of QD1 spin $z$-component
flips at the transition. Notice, that should the $t'$ sign happen to change, 
the exchange field and all the polarizations would change the sign as well 
(as long as the trimer is at local PHS point).

Furthermore, the side-coupled quantum dots, QD2 and QD3, are actually even stronger polarized, see \fig{Sz2D} 
and compare \fig{Sz1D}(a) with \fig{Sz1D}(b). 
In fact, in the UK phase and for $T< \ex^{\rm GS}$ they are completely polarized with 
$\Spin_{z2}=\Spin_{z3}=-1/2$ ($\Spin_{z2}=\Spin_{z3}$ is a consequence of symmetry and further we only discuss
$\Spin_{z2}$). The value in the FK phase is, on the other hand, $\Spin_{z2} \approx -1/3$ and decreases slowly 
with increasing $t$ to obtain $\Spin_{z2} = 1/4$ for $t \gg U$, which is still significantly larger than for QD1
and causes significant net trimer polarization, $\Spin_z\approx -1/2$. 
This kind of magnetic ordering is quite surprising in the Kondo regime, especially at the local PHS point
and for very weak values of frustrating coupling $t'$. It is intriguing, if the realization of a similar state
is possible in some correlated frustrated lattice. Moreover, it is also intriguing
whether the realization of a similar state may be possible in correlated,
frustrated lattices.

\subsection{Trimer entropy}
\label{sec:entropy}

\begin{figure}[t]
\centering
\includegraphics[width=1\linewidth]{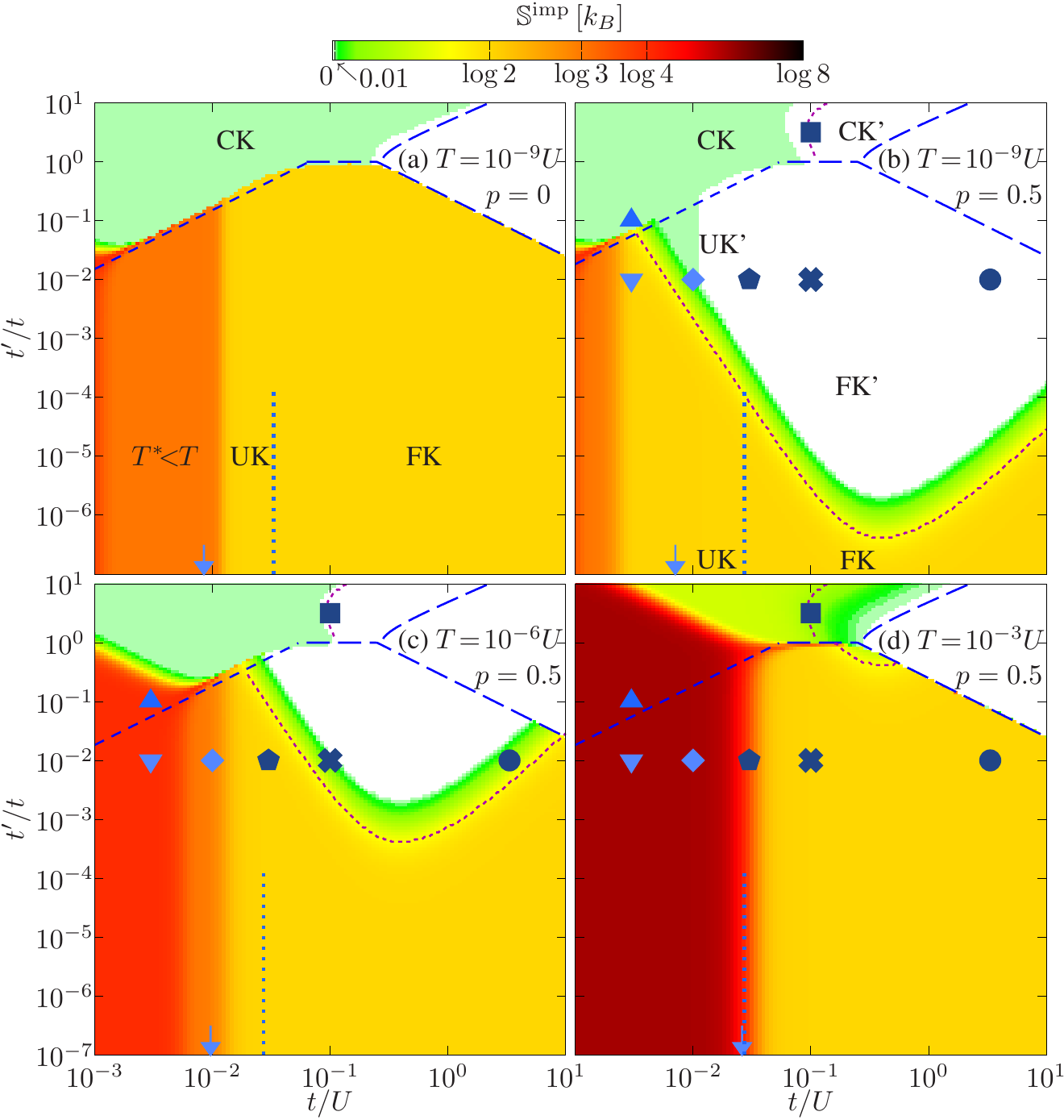}
\caption{
		 The entropy of the trimer $\Simp$ as a function of $t$ and $t'$. Note that the white 
		 regions correspond to $\Simp \ll 0.01 k_B$. The parameters are the same as in \fig{G2D}.
		 The points marked with symbols indicate the values of $t$ and $t'$ for which \fig{S1D} is prepared.
		 Small vertical arrows (dotted lines) indicate $t^*$ ($t_x$) positions, correspondingly. 
		 } 
\label{fig:entropy2D}
\end{figure}

As a confirmation of the interpretations derived in the preceding sections, 
in the following we present the results concerning the trimer entropy $\Simp$,
defined in \eq{SimpDef}. The simple intuition is that $\Simp/k_B$ is a logarithm 
of the ground state degeneracy at $T=0$, and it is increased by the number 
of trimer states available by thermal fluctuations at elevated temperatures.
In \fig{entropy2D}(a) we present the results for $p=0$, which extend the range of validity of earlier results by Mitchell
{\it et al.} \cite{Mitchell_QPT,Mitchell_phase_diag}. In the CK phase, where 
the ground state is the conventional Kondo singlet, $\Simp$ practically 
equals $0$. More precisely, it has a small value $\Simp \approx 0.01 k_B$
due to finite temperature used for calculations, $T=10^{-9}U$. It is even 
more strongly suppressed in the $\Spin = 0$ regime, where no small energy scale 
(such as $T_K$) is relevant. Then, on the other side of QPT, where 
the ground state is the spin doublet, the trimer entropy $\Simp/k_B = \log 2$, 
as should be expected. Moreover, exactly at the transition the degeneracy 
between the singlet and triplet states gives rise to $\Simp/k_B = \log(3)$ (at nonzero 
temperatures this remains true within a small vicinity of the transition).

Due to finite temperature, this result is also valid in the UK phase 
for $t<t^*$, when the spin triplet formed within the QD2-QD3 subsystem is 
not yet screened; compare the position of the crossover between
the values of $k_B \log(2)$ and $k_B \log(3)$ marked with 
the vertical arrow corresponding to $t=t^*$ in \fig{entropy2D}(a). 
Obviously, at the transition point between this phase and the CK phase one finds
$\Simp/k_B = \log(4)$. However, since in the FK phase the ground state degeneracy 
equals that of the UK phase for $t>t^*$, the FK/UK crossover (its position is 
indicated in the figure with a dashed line) does not give rise to any signature 
in the value of $\Simp$, as opposed to the spin expectation value; cf. \fig{S2D}(a).

The landscape changes significantly, when the leads magnetization is taken into 
account. Then, the effective exchange field induced in the trimer, $\ex$,
splits the spin multiplets, lifting degeneracy of the ground state. Therefore,
whenever $\ex > k_B T$, the trimer entropy drops to zero, manifesting the 
crossover from magnetic to non-magnetic state. This is clearly visible in
the FK and FK' phases in \fig{entropy2D}(b), as well as for the UK and UK' phases therein.
Similarly, while in the non-magnetic CK phase $\Simp\approx 0.01 k_B$, 
in the case of magnetic leads and the corresponding CK' regime
the Kondo effect is suppressed when $\ex> T $
and so is the trimer's entropy; cf. \fig{entropy2D}(b). 
Nevertheless, this makes the value of $\Simp$ the same in all magnetic 
phases, and the QPT is visible only as a peak of height $k_B \log(2)$ 
in the region of QPT-related degeneracy.


All the above results remain generally correct for higher temperatures, 
see Figs.~\ref{fig:entropy2D}(c)-(d), however, phase borders are slightly shifted
and regions corresponding to degeneracy at QPT are broadened. Moreover, 
for sufficiently small values of $t$ and $t'$, the excited states become available,
which results in an increase of the trimer's entropy.

\subsection{Other quantities}
\label{sec:quantities}

We found the set of quantities analyzed in previous sections, namely the linear conductance $G$, 
the trimer spin and its $z$-component expectation values, $\Spin$ and $\Spin_z$, 
and the trimer entropy $\Simp$, sufficient for complete determination of all
the relevant phases. However, in general,
the analysis of other physical quantities may also be helpful to pinpoint 
all the regimes in complex systems. In particular, some of the magnetic 
regimes of the system studied here may be recognized in the maps of 
tunneling magneto-resistance or current spin polarization. However, the
characteristic features are restricted to a few of many regimes of the 
phase diagram and lack some general explanations allowing to hope that
similar features may be relevant for different structures. On the other hand, thermoelectric 
quantities, such as the Seebeck coefficient, can help to determine the phases
in strongly particle-hole asymmetric systems, where they tend to have 
large values and alternating sign in different regimes \cite{CostiZlatic}.
The dynamical spin-spin susceptibilities occur to be especially useful
for analysis of spin-symmetric non-Fermi-liquid phases \cite{Wang2020}. 
The exceptional usefulness of the quantities chosen here, in particular 
$\Spin$ and $\Spin_z$, stems from the fact, that they
characterize equally well non-magnetic as well as fully or partially magnetized 
structures, which is crucial for determination of the local magnetic 
texture. This feature may prove important for theoretical characterization 
of the bulk materials possessing the same correlations as the system 
described here, in particular frustrated heavy-fermion materials.

\section{Conclusions}
\label{sec:conclusions}

We have determined and analyzed the properties of strongly-correlated frustrated quantum dot trimer coupled to ferromagnetic leads. 
The considerations have been performed by using the numerical renormalization group method,
which was used to calculate the conductance, spin expectation values and the entropy
of the analyzed quantum dot trimer. 
This allowed us to construct the full phase diagram of the system as a function
of hoppings between the dots and the temperature, together with the corresponding phase boundaries.
We showed that as the hoppings are tuned, at $T=0$ and for nonmagnetic leads,
the system can reveal different phases:
the conventional Kondo phase, the underscreened Kondo phase,
the ferromagnetic Kondo phase as well as the non-Kondo spin-less phase.
These phases are present at finite temperatures,
but are not stable in the limit of vanishing temperature in the presence 
of arbitrarily weak frustrating coupling. 
We then determined the fate of different Kondo phases in the case of ferromagnetic leads,
when the spin splitting of states occurs due to an exchange field.
Interestingly, such exchange field can be generated even at the local
particle-hole symmetry point of each dot if the frustrating hopping,
which breaks the global PHS, is finite.
We showed that the spin polarization of the trimer in the Kondo regime may persist 
up to sizable temperatures even when the frustrating coupling is very small.
This allowed us to extend our conclusions to molecular trimers effectively coupled to one conduction channel, 
where the frustration is introduced by next-nearest-neighbor hopping.
Potentially, these results may be also of relevance for frustrated correlated lattices, 
where the Kondo screening may coexist with magnetic ordering, if some of the local moments are
coupled to the electronic bath only indirectly.

\begin{acknowledgements}
This work was supported by the National Science Centre in 
Poland through the projects No.~2015/19/N/ST3/01030 (KW)
and No.~2017/27/B/ST3/00621 (IW)
and by the Deutsche Forschungsgemeinschaft (DFG) through the Cluster of
Excellence ML4Q (390534769) (J.K.).
K.W. also acknowledges the fellowship of the Alexander von Humboldt Foundation. 
\end{acknowledgements}


%

\end{document}